\documentstyle[namedreferences]{kluwer}
\runningtitle{Heat and Kinetic Theory in 19th-Century 
Textbooks}
\runningauthor{J. M. VAQUERO AND A. SANTOS}

\begin{opening}

\title{Heat and Kinetic Theory in 19th-Century Physics
Textbooks: The 
Case of Spain}

\author{JOS\'{E} M. \surname{VAQUERO}}
\author{ANDR\'{E}S \surname{SANTOS}}
\institute{Departamento de F\'{\i}sica, Universidad de Extremadura, E-06071 
Badajoz, Spain} 
\end{opening}

\begin{document}

\begin{abstract}
Spain was a scientifically backward country in the early 19th-century.  The 
causes were various political events, the War of Independence, and the reign 
of Fernando VII. The introduction of contemporary physics into textbooks was 
therefore a slow process. An analysis of the contents of 19th-century 
Spanish textbooks is here presented, centred on imponderable fluids, the 
concept of energy, the mechanical theory of heat, and the kinetic theory of 
gases.
\end{abstract}

  



\section{INTRODUCTION}
The period of Enlightenment was a time when Spanish physics,  which had been 
lagging behind the level of the rest of Europe, was able to recover. 
However, the political events of the end of the 18th-century, the War of 
Independence, and then the reign of Fernando VII ruined the panorama of 
Spanish science. History was sadly to repeat itself with the Spanish Civil 
War of 1936, which brought to nothing the efforts of physicists and 
mathematicians of the end of the 19th and the beginning of the 20th 
centuries to regain the time lost during the reign of Fernando VII until the 
``generation of '98''.

Part of the disastrous situation during the 19th-century can be traced 
back to the legislation and curricular plans of Spain's universities. The 
Faculty of Exact, Physical, and Natural Sciences was not created until 1857, 
with the Ley (Law) Moyano. Physics had been relegated to a minor faculty, 
that of Arts, as preparation for the major faculties, and formed part of the 
``Philosophical Institutions'' which were studied in the following order: I. 
History of philosophy and elements of mathematics; II. Logic and 
metaphysics; III. General physics; IV. Special physics. All was in a Latin 
which had become progressively less suited to teaching, so that enlightened 
reformers always attempted to publish textbooks in Spanish. Physics texts in 
Latin were still being imposed on students even up to the time of the 
absolutist period of Fernando VII \cite{Mor88}.

After the death of Fernando VII, an ordinance regulating printing  was 
promulgated in 1834 which allowed a certain freedom in publication, in 
particular in scientific publishing. One of the consequences of the 
centralization of curricular plans, such as the Plan Pidal, was the impulse 
given to the production of textbooks associated with the new programs of 
study. They were original productions as well as translations of foreign 
authors, and formed the beginning of a national scientific output which was 
to be of greater or lesser quality according to each particular case. The 
study plans usually counselled the use of certain textbooks which most 
closely approached the spirit of the courses. The procedure consisted in 
appointing committees to decide on the most suitable texts which then put 
forward a number of them to be chosen from.

In the meantime, Europe was seeing a major change in vision towards  atomism 
thanks to such personalities as Herapath, Waterston, Clausius, van der 
Waals, Maxwell, and Boltzmann \cite{Bru86}. Imponderable fluids had been 
abandoned as the explanation of physical phenomena, and Mechanics, Heat, 
Electromagnetism, and Optics had been unified by the concept of energy. 
Teaching was based on a program of mechanical explanations \cite{Har82}.

The goal of the present communication is to analyze the impact of modern  
ideas concerning energy and the constitution of matter on the textbooks of 
19th-century Spain. We have examined 45 works on general physics at 
secondary and university education levels, with publication dates from the 
mid 19th-century to the early 20th-century. One of the first things that 
struck us was the great similarity between the books of different authors. 
With centralism, teachers were encouraged to write their own textbooks and, 
indeed, were rewarded for doing so. At the same time, the administration's 
goal was for teaching to be uniform nationwide. The result was that 
textbooks were written essentially with the Ministerio de Fomento's 
(Ministry of Development) official program as the Table of Contents, and 
that the texts served solely to expound known science and in no way to serve 
as the basis for further research. There were those who, being interested in 
science teaching, protested about the policy concerning textbooks. 
Representative of them was Eduardo Lozano y Ponce de Le\'{o}n. 
Under the pseudonym L. Opando y Uceda, Lozano published ``Programas y Libros 
de Texto'' (Programs and Textbooks) in {\em Revista de la Sociedad de 
Profesores 
de Ciencias} \cite{Opa75}, and under the pseudonym ``Un Extreme\~no'' 
the same article in the journal {\em El Magisterio Extreme\~no} ({\em The 
Extreme\~nan 
Schoolmaster}) \cite{Ext75}. In this same journal and on the same theme, 
Ildefonso Fern\'andez S\'anchez published another critical article 
\cite{Fer76}. In his writings Lozano asked for the programs to have 
reference strictly to the subject matter to be dealt with and not to the 
methods to be used, and that the programs should include suggestions of the 
books that were best suited to the subject, but without the obligation of 
following them.

The immediate consequence of that educational policy was that the different  
editions of textbooks seemed rather to be reprintings: textbook 
``immutability'' from one edition to another was astonishing. Let us take two 
cases as example. The {\em Trait\'{e} \'{E}l\'{e}mentaire de Physique} ({\em 
Elementary 
Treatise on 
Physics}) by A. Ganot was used in many Spanish 
and European Institutes (secondary education) and Universities. The first 
Spanish language edition dates from 1853. The 18th Spanish edition was 
printed in 1923, and even as late as 1945 an edition was published in San 
Sebasti\'an \cite{Mor88}. In all this time there were only small 
modifications made with respect to the first French edition. An even more 
extreme example, since it was a work with even more antiquated ways of 
putting the material than Ganot, was the textbook of \citeauthor{Gon51}. 
The 2nd edition dates from 1851 \cite{Gon51} 
and the 10th edition from 1870 \cite{Gon70}. 
In these twenty years and eight new editions the work remained without a 
single significant change, notwithstanding the archaic ways of setting forth 
the subject matter of the first editions. A third example is less 
disappointing: {\em Elementos de F\'{\i}sica} ({\em Elements of Physics}) by 
Enrique Iglesias 
Ejarque. The first edition of 1897 \cite{Igl97} has quite a modern manner 
of exposition. Nevertheless, while the following editions introduce small 
amounts of additional material into the text, there are never any 
significant changes. The 8th edition dates from 1924 \cite{Igl24} and the 
last that we can find a reference to is the 10th edition in 1933. Clearly 
one may conclude that physics textbooks had quite a long effective lifetime.

While there was a certain degree of difficulty in introducing new material  
into the textbooks, the great problem was to eliminate content which was 
included by tradition even though it was antiquated. The case of Eduardo 
S\'anchez Pardo, the translator of Ganot's work, is significant. In the 
prologue to that popularly used book:

\begin{quote}
{\small The editor D. C\'arlos Bailly-Bailli\`{e}re being for his part 
desirous 
that 
 our public in general, and the pupils of our Institutes and Faculties in 
particular, follow science in her latest advances, charged us with the 
translation of the latest edition of the cited Treatise on Physics. On 
taking on this commitment, with the object that the Spanish edition be more 
complete, we judged it convenient, corresponding thereby also to the wishes 
of the editor, to conserve certain theories, the exposition of various 
experiments, and the description of some of the instruments or apparatus 
that had figured in earlier editions of the forementioned work, and that in 
the latest edition had been totally or partially suppressed.
(\citeauthor{Gan76} \citeyear{Gan76}, p. v)}
\endnote{The original text is:
        ``Deseoso por su parte el editor D. C\'arlos Bailly-Bailli\`{e}re de 
que 
nuestro p\'{u}blico en general, y particularmente los alumnos de nuestros 
Institutos y Facultades sigan \'a la ciencia en sus \'{u}ltimos adelantos, nos 
encomend\'{o} la traduccion de la \'{u}ltima edicion del citado Tratado de 
F\'{\i}sica. Mas al encargarnos de este cometido, con objeto de que la edicion 
espa\~nola fuera mas completa, juzgamos conveniente, correspondiendo 
por otra parte as\'{\i} \'a los deseos del editor, conservar algunas 
teor\'{\i}as, la 
exposicion de varios experimentos y la descripcion de algunos 
instrumentos \'{o} aparatos que en ediciones anteriores de la mencionada 
obra figuraban, y en la \'{u}ltima h\'anse total \'{o} parcialmente 
suprimido.''}
\endnote{Among the old material that was revived one can find, for instance, 
sections on Animal Electricity,  Perreaux's dynamometer, Alvergnat's 
barometer, Carr\'e's device for making ice or Sturm's vision theory.}
\end{quote}

 Another of the characteristics of Spanish physics textbooks was their  
orientation to student success in tests and examinations. Works that are 
halfway between textbook and simple program of the curriculum abound. 
Secondary education teachers often encouraged the students to think little 
about the physical phenomena themselves but to learn definitions, laws, and 
descriptions of apparatus. As clear examples we could cite, amongst several 
works, the {\em Resumen de F\'{\i}sica y Nociones de Qu\'{\i}mica} ({\em A 
Summary of 
Physics and 
Notions of Chemistry}) \cite{San65}, and {\em Definiciones, Principios 
y Leyes de la F\'{\i}sica} ({\em Definitions, Principles, and Laws of 
Physics}) 
\cite{Paz92}.
 Hence, for most students learning was not meaningful but 
purely memorization. Together with the lack of practical classes, this may 
have been one of the causes of Spanish physics' slow rate of development 
during the 19th-century.
\section{ANALYSIS OF THE TEXTBOOKS}

In order to analyze what we understand to have been the introduction of  
modern physics into Spanish textbooks, we asked ourselves the following 
questions relative to the works that we consulted:
\begin{enumerate}
\item 
\label{q1}Are imponderable fluids studied in the textbook?
\item \label{q2} Is the term caloric used to refer to heat?
\item \label{q3}Does the concept of energy appear in a general form?
\item \label{q4}Does the mechanical theory of heat appear?
\item \label{q5}Does the kinetic theory of gases appear?
\end{enumerate}
Affirmative responses to the first two questions would indicate  
traditionalism, and to the last three, modernity. We would classify a 
textbook as modern if questions \ref{q3} and \ref{q4} are answered 
affirmatively. If the kinetic theory of gases appears as well (question 5), 
then we would consider the textbook to be quite complete. Table \ref{TableI} 
lists the responses to these questions for the 19th-century textbooks to 
which we had 
access. The works are ordered by publication date. When the edition is not 
given, reference is to the first edition.

Most of the works consulted belong to the library of the Real Sociedad 
Econ\'{o}mica de Amigos del Pa\'{\i}s in Badajoz (Spain). While not 
exhaustive, 
the sample is significant, since students of the city of Badajoz in the 
19th-century were prepared in this centre for the entrance examinations of 
the Schools of Engineering (high-level technical degree courses, rather than 
purely Engineering in the English language sense) or Military Academies.

\begin{table}
\caption{
Responses to the five questions posed in the text, according to the 
different  works consulted.}
\label{TableI}
\begin{tabular}{lccccc}\hline
Textbook & \ref{q1}  & \ref{q2}&\ref{q3}&\ref{q4}&\ref{q5}\\ \hline         
\cite{Rib44}&Yes&Yes&No&No&No\\
\cite{Deg45}, 2nd ed.&No&No&No&No&No\\
\cite{Mor45}&Yes&Yes&No&No&No\\
\cite{San46}&Yes&Yes&No&No&No\\
\cite{Pin47}&Yes&Yes&No&No&No\\
\cite{Gon51}, 2nd ed.&Yes&No&No&No&No\\
\cite{Gon56}, 4th ed.&Yes&No&No&No&No\\
\cite{Gon57}, 5th ed.&Yes&No&No&No&No\\
\cite{Rod58}&Yes&Yes&No&Yes&No\\
\cite{Fer61}, 2nd ed.&Yes&Yes&No&No&No\\
\cite{San65}&Yes&Yes&No&Yes&No\\
\cite{Bou66}&No&No&No&Yes&No\\
\cite{Gon68}, 9th ed.&Yes&No&No&No&No\\
\cite{Ric69}, 7th ed.&Yes&Yes&No&No&No\\
\cite{Gon70}, 10th ed.&Yes&No&No&No&No\\
\cite{Fel74}, 2nd ed.&Yes&No&No&No&No\\
\cite{Ric75}, 8th ed.&Yes&Yes&No&No&No\\
\cite{Gan76}, 7th ed.&No&No&No&Yes&Yes\\
\cite{Fue79}&No&No&No&Yes&No\\
\cite{Ram80}, 6th ed.&Yes&Yes&No&Yes&No\\
\cite{Fue82}, 2nd ed.&No&No&No&Yes&No\\
\cite{Ric82}, 10th ed.&Yes&Yes&No&No&No\\
\cite{Mar86}&No&No&Yes&Yes&No\\
\cite{Pin87}&No&No&No&Yes&No\\
\cite{Ami89}&No&No&Yes&Yes&Yes\\
\cite{Pic89}&No&No&No&No&No\\
\cite{Fel90}, 7th ed.&No&No&Yes&Yes&No\\
\cite{Esc91}&No&No&Yes&Yes&No\\
\cite{Paz92}&No&No&Yes&Yes&Yes\\
\cite{Gar92}&No&Yes&Yes&No&No\\
\cite{Loz93}, 3rd ed.&No&No&Yes&Yes&No\\
\cite{Mar93}&No&No&Yes&Yes&No\\
\cite{Rod95}, 2nd ed.&Yes&Yes&Yes&Yes&No\\
\cite{Rib95}&No&Yes&Yes&Yes&No\\
\cite{Fel96}, 8th ed.&No&No&Yes&Yes&No\\
\cite{Igl97}&No&No&Yes&Yes&Yes\\
\cite{Sol00}, 2nd ed.&No&No&Yes&Yes&Yes\\
\cite{Loz00}&No&No&Yes&Yes&No\\
\hline
\end{tabular}
\end{table}

Some comments are necessary concerning Table \ref{TableI}. The books that 
were published in Spain in the 1840s studied imponderable fluids, and dealt 
neither with energy, nor with the mechanical theory of heat, nor with the 
kinetic theory of gases. This situation was natural since these concepts 
were only being developed in Europe at this time. However some foreign 
authors already did not study imponderable fluids (neither did they use the 
term caloric to refer to heat), and their Spanish translations set an 
example for such fluids to be abandoned. This was the case with the textbook 
of Deguin, translated by Venancio Gonz\'alez \cite{Deg45}. The translator 
would write his own textbook later \cite{Gon51}, and despite 
going through numerous editions, would never abandon imponderable fluids. 
Another textbook of a French author is \citeauthor{Pin47} \shortcite{Pin47}, 
translated by 
Florencio Mart\'{\i}n  Castro, although here imponderable fluids were still 
being studied in the original.

The handbook of \citeauthor{Mor45} \shortcite{Mor45} presents physics as a  
science of Nature in general, including subjects such as geology and 
geography. This was the concept of physics that existed in Europe at the 
beginning of the century. On the positive side, the mechanical ideal which 
was at the base of the development of physics in the 19th-century did 
indeed have a reflection in Spanish textbooks. One reads in the prologue of 
the work of \citeauthor{Rib44} \shortcite{Rib44}:

\begin{quote}
{\small (\ldots) Effectively, all the phenomena attributed to caloric, to 
light, and to  the electric fluid are mensurable and calculable effects; all 
are attributed to forces, all consist of movements, and constitute, in sum, 
mechanics.
(\citeauthor{Rib44} \citeyear{Rib44}, p. i})\endnote{The original text is:
``(\ldots) Efectivamente, todos los fen\'{o}menos atribuidos al cal\'{o}rico, 
\'a 
la luz y al fluido el\'{e}ctrico, son efectos mensurables y calculables; todos 
se atribuyen \'a fuerzas, consisten todos en movimientos, y constituyen en 
fin la mec\'anica.''}
\end{quote}

The texts that we were able to consult from the 1850s are mainly editions 
of the work of \citeauthor{Gon51} 
(\citeauthor{Gon51} \citeyear{Gon51}, \citeyear{Gon56}, \citeyear{Gon57}).
The principal characteristic of this
text was its immutability, edition after edition. Although it studied 
imponderable fluids, the term caloric was not used, and neither energy, nor 
the mechanical theory of heat, nor the kinetics of gases appear in the text. 
The other textbook that we consulted is that of \citeauthor{Rod58} 
\shortcite{Rod58}. 
Although imponderable fluids are studied and the term caloric is used, the 
author briefly explains the theory of ondulations 
\endnote{According to the hypothesis of ondulations, as defined in the 
19th-century Spanish books we have consulted, heat is caused by the rapid 
motion of the molecules and is transmitted through the aether by ondulations.
Thus, all the heat phenomena are referred to a unique cause, motion, in 
contrast to the heat understood as a substance (caloric).
The hypothesis of ondulations is not exactly the same as the wave theory of 
heat \cite{Bru86}, according to which heat is the vibrations of aether 
itself.}
(as the mechanical theory 
of heat was first known), and says of that theory that ``it is the one which 
today seems more correct'' (p. 173). This work was awarded a prize in a 
public competition under the auspices of the Real Academia de Ciencias 
(Gaceta de Madrid, 9 September 1854). Neither the concept of energy in a 
general form nor the kinetic theory of gases appear in Rodr\'{\i}guez's text. 
It 
is interesting too that in this decade there appeared a physics textbook 
written in Latin which followed the scholastic tradition \cite{San57}.

The textbooks of Spanish authors of the 1860s were still studying  
imponderable fluids. They were also still using the term caloric to refer to 
heat, except for the editions of \citeauthor{Gon68} (\citeyear{Gon68},
\citeyear{Gon70}),
as we noted before.
One French-authored textbook, \citeauthor{Bou66} \shortcite{Bou66}, did not 
study 
imponderable fluids or use the term caloric, but then neither did it use the 
energy concept or the theories we are looking for. The same was the case 
with the textbook of \citeauthor{Deg45} \shortcite{Deg45}, commented on above. 
A possible reason 
for these to be missing is that the translator, Ram\'{o}n de la Sagra, used 
the 
7th French edition.

The note of modernity is found in the textbook of \citeauthor{San65} 
\shortcite{San65}, in which there appears an idea concerning the 
mechanical theory of heat, namely the hypothesis of 
ondulations, although for didactic purposes the theory of emissions 
(caloric) was preferred. Thus, one reads:

\begin{quote}
{\small The system of ondulations is the most scientific, and the most 
admitted in  modern physics; but that of emission lends itself more to 
demonstrations, for which reason it is generally preferred for the 
explanation of the phenomena of the caloric. (\citeauthor{San65} 
\citeyear{San65}, p. 
204)}\endnote{The
original text is:
        ``El sistema de las ondulaciones es el mas cient\'{\i}fico, y el mas 
admitido en la f\'{\i}sica moderna; pero el de la emisi\'{o}n se presta m\'as 
a las 
demostraciones, por lo que se prefiere generalmente para la esplicacion 
de los fen\'{o}menos del cal\'{o}rico.''}
\end{quote}

The textbooks of the 1870s begin to abandon the traditional theses, and  
show signs of modernity. The most traditional are \citeauthor{Ric75} 
\shortcite{Ric75},
an 8th edition, and \citeauthor{Fel74} \shortcite{Fel74}. The latter already 
does not 
use the term caloric, and later editions were progressively modernized.

The textbook of \citeauthor{Ram80} \shortcite{Ram80} studies imponderable 
fluids, and uses  
the term caloric. In its treatment of radiant heat, however, the work seems 
very modern. As noted above for \citeauthor{San65} \shortcite{San65}, the 
author prefers 
the caloric hypothesis for its simplicity in teaching:

\begin{quote}
{\small The admissible hypothesis is at present that of ondulations, in the 
light  of the advances in modern physics; but as it simplifies the 
demonstrations, many physicists prefer the hypothesis of emission to explain 
the phenomena of heat. (\citeauthor{Ram80} \citeyear{Ram80}, p. 
148)}\endnote{The 
original text is:
        ``La hip\'{o}tesis de las ondulaciones es         la 
admisible en la actualidad, atendidos los progresos de la f\'{\i}sica moderna; 
pero simplific\'andose las demostraciones por la hip\'{o}tesis de la 
emisi\'{o}n,
 muchos f\'{\i}sicos la prefieren para esplicar los fen\'{o}menos del 
calor.''}
\end{quote}

The textbooks with a more modern spirit are those of \citeauthor{Fue79} 
\shortcite{Fue79}  
and \citeauthor{Gan76} \shortcite{Gan76}. M\'aximo Fuertes Acevedo's work 
neither studies imponderable fluids nor uses the term caloric. But energy 
does not appear as a general concept either, despite the mechanical theory 
of heat being explained. Adolphe Ganot's textbook was much used in Europe. 
The first Spanish edition is of the year 1853. We consulted the 7th Spanish 
edition \cite{Gan76}. It contains a paragraph explaining the dynamic theory 
of gases: gases are described as formed by elastic molecules in motion, and 
the elasticity of a gas at a given volume is proportional to the {\em vis 
viva} (total mass of the molecules multiplied by the square of their 
speed). While making use of such concepts as this (today, of course, 
replaced by kinetic energy), the text does not deal with energy in a general 
form. As is to be expected with this perspective, imponderable fluids and 
the term caloric have been forgotten, and the mechanical theory of heat is 
studied.

All the textbooks of the 1880s that we consulted have abandoned 
imponderable  fluids, with the exception of the handbook of \citeauthor{Ric82} 
\shortcite{Ric82} 
which is already in its 10th edition. Of the other works, 
the most traditional is that of \citeauthor{Pic89} \shortcite{Pic89}, since it 
neither deals 
with energy in a general way nor introduces the theories we are looking for. 
We then have the 2nd edition of the work of \citeauthor{Fue82} 
\shortcite{Fue82},
with no substantial changes from the 1st edition.
We also consulted a Portuguese secondary education textbook,  
\citeauthor{Pin87} \shortcite{Pin87}, which has a similar perspective to 
those of \citeauthor{Fue79} (\citeyear{Fue79}, \citeyear{Fue82}). The textbook 
of 
\citeauthor{Mar86} \shortcite{Mar86} is the first of the series of books 
that we consulted in which energy is dealt with in a general fashion. The 
mechanical theory of heat is studied, but the kinetic theory of gases has 
still not appeared. The 7th edition of the work of \citeauthor{Fel90} 
\shortcite{Fel90} has almost nothing to do with the 2nd edition that we 
commented on above \cite{Fel74}. Now, energy and the 
mechanical theory of heat are presented. In the prologue to the 6th edition, 
also included in the 7th, one reads:

\begin{quote}
{\small With great insistence I have attempted in the treatment of heat to 
relate  together all the phenomena of thermo-dynamic theory.
(\citeauthor{Fel90} \citeyear{Fel90}, p. v)}\endnote{
The original text is:
        ``Con insistencia grande he procurado en el tratado del calor hacer 
relacionar todos los fen\'{o}menos con la teor\'{\i}a termo-din\'amica.''}
\end{quote}

The outstanding textbook of this decade is that written by   
\citeauthor{Ami89} \shortcite{Ami89}, published in Tarragona. The author held 
the chair of 
physics and chemistry in the Instituto Provincial of Tarragona. This is a 
modern text which includes the kinetic theory of gases, and in general 
explains physical phenomena mechanistically. In the work's prologue, the 
author speaks about a book of his on mechanics published in 1885, and gives 
great importance to this branch of physics:

\begin{quote}
{\small The criterion that has inspired this treatise responds to the 
necessity  already recognized by all to explain the subject of Physics in a 
single course, always preceded by a short course of Mechanics as foundation 
and basis of the former (\ldots). (\citeauthor{Ami89} \citeyear{Ami89}, p. 3)}
\endnote{The
original text is:
        ``El criterio que ha inspirado este tratado, responde \'a la 
        necesidad 
ya reconocida por todos, de explicar en un s\'{o}lo curso la asignatura de 
F\'{\i}sica, precedida siempre de un cursillo de Mec\'anica como fundamento y 
base de aquella (\ldots)''}
\end{quote}

As a continuation of thermodynamics, \citeauthor{Ami89} describes the kinetic 
theory of  
gases by following the ideas of Clausius and introducing the definition of 
free path of a molecule as the distance travelled between two consecutive 
collisions. He deduces Mariotte's Law, obtaining the formula
$p V=\frac{1}{3}nmv^2$, where $p$ is the pressure of the gas, $V$ its 
volume, $n$ the number of molecules, $v$ the mean velocity, and $m$ the mass 
of a molecule. He also deduces Avogadro's hypothesis from the kinetic theory 
of gases \cite{Vaq98}. As an indication of the little that this 
personality and his work have been studied, there only appears one book of 
\citeauthor{Ami92} in the {\em Collected Catalogue of Spain's Bibliographic 
Heritage: 19th Century} \cite{Bib89}. This is a 
textbook on elementary chemistry \cite{Ami92}.

The main characteristic of the books of the 1890s is that they all now  
treat energy in a general form, as well as the mechanical theory of heat. 
The textbook of \citeauthor{Rod95} \shortcite{Rod95} simultaneously uses 
imponderable 
fluids and the term caloric, but only as additional information. A more 
curious case is that of \citeauthor{Rib95} \shortcite{Rib95} whose textbook 
uses 
the term caloric without studying imponderable fluids.

The work of \citeauthor{Esc91} \shortcite{Esc91} has the interest of 
studying  heat 
and light side by side
\endnote{On the other hand, \citeauthor{Esc91} does not distinguish radiant 
heat from heat as energy of molecular motion}, to bring out their 
relationship as vibratory phenomena:

\begin{quote}
{\small (\ldots) The molecular vibrations of the bodies, 
transmitted by the aether, produce the feelings of HEAT in the 
touch and of LIGHT in the sight. 
 (\citeauthor{Esc91} \citeyear{Esc91},
p. 496)}\endnote{The original
text is:
        ``(\ldots) Las vibraciones moleculares de los cuerpos, transmitidas 
        por el \'eter, producen en el tacto la sensaci\'on de CALOR y en la 
        vista de LUZ.''}
        
\end{quote}

 Other modern textbooks are those of 
\citeauthor{Loz93} \shortcite{Loz93},
the pharmacist \citeauthor{Gar92} \shortcite{Gar92}, \citeauthor{Mar93} 
\shortcite{Mar93},
and \citeauthor{Fel96} \shortcite{Fel96}.

The small book of  \citeauthor{Paz92} \shortcite{Paz92} contains only 
definitions and  
principles. At no time is any idea developed, since the book's object is to 
serve as a collection of phrases for students to learn in preparation for 
their examinations. As part of these examination aids, the author prepared 
two plates which accompany the text, one on units and abbreviations of the 
decimal metric system, and another on physical units. The kinetic theory of 
gases does not appear explicitly, but some of its results do. For instance, 
one can read Maxwell's Law:

\begin{quote}
{\small The viscosity of a gas measured by the coefficient of friction is  
independent of the density. (\citeauthor{Paz92} \citeyear{Paz92},
p. 60)}\endnote{The original
text is:
        ``La viscosidad de un gas medida por el coeficiente de frotamiento 
es independiente de la densidad.''}
\end{quote}

Another modern work is that of \citeauthor{Igl97} \shortcite{Igl97}. The 
author  
indicates on the first page in a footnote that the works that had been 
consulted were ``Spanish: Escriche, Feli\'{u}, Mu\~noz, Rojas, and 
Rodr\'{\i}guez Largo. 
Foreign: Ganot, Jamin, Joubert, Maxwell, and Tyndall''. The text contains a 
short paragraph on the Theory of Gases, in which Bernouilli, Clausius, and 
Maxwell are cited. A kinetic interpretation of pressure is also given, but 
neither is the concept of mean free path introduced nor estimates of 
molecular speeds given.

The textbook of \citeauthor{Sol00} \shortcite{Sol00}, another modern work, has 
in the  
book dedicated to heat an article on the {\em Thermal Constitution of Gases}, 
in 
which the kinetic theory is described. A kinetic interpretation of pressure 
is given, together with an explanation of high molecular velocities, 
including numerical estimates. A short section is dedicated to the {\em Height 
of 
the Atmosphere}, and another to the {\em Mean Free Path}, in which one reads 
that 
``Crookes calls mean free path the space travelled by the molecule between 
two of those collisions''.
On the other hand,  the kinetic theory of gases is not mentioned by
\citeauthor{Loz00} \shortcite{Loz00}. 

We also consulted some 20th-century books. They all had a modern 
perspective. 
 The textbooks of \citeauthor{Car25} \shortcite{Car25} and 
 \citeauthor{Mon28} \shortcite{Mon28}
 did not include the kinetic theory of gases. The 6th and 8th 
 editions of \citeauthor{Igl15} (\citeyear{Igl15}, \citeyear{Igl24}) present 
improvements over the 1st 
 edition. With respect to the kinetic theory of gases, a new paragraph is 
 included to explain molecular velocities. The textbook of \citeauthor{Gon04} 
\shortcite{Gon04}
 explains van der Waals's equation of state.

The last textbook that we wish to comment on is a translation of an Italian  
book. It is the work of \citeauthor{Cas32} \shortcite{Cas32}, a university 
textbook of 
modern physics where the kinetic theory of gases is developed completely. The 
topics in the book are surprising in their breadth and modernity.

As a qualitative summary, Figure \ref{fig1} shows the evolution of the 
responses to 
 the questions that we formulated, according to the data listed in Table 
 \ref{TableI}.
\begin{figure}
\vspace{8cm}
\caption{
The evolution of textbook contents with respect to the abandoning 
of imponderable fluids and the introduction of the concept of energy, the 
mechanical theory of heat, and the kinetic theory of gases. The figure has 
been constructed by calculating for each decade the percentage of 
affirmative responses to the questions formulated in the text. The intention 
of these results is not to provide full statistical certainty, but to 
present qualitatively the findings of this investigation. One clearly 
appreciates the decline in the use of imponderable fluids and the rise of 
content related to an atomic view of matter.}
\label{fig1}
\end{figure}

\section{CONCLUSIONS}
Spain's educational policies in the 19th-century encouraged a great  
similarity between the general physics textbooks of different authors, 
although there were voices such as that of Eduardo Lozano y 
Ponce de Le\'{o}n, in discord with this policy. This situation led to the 
differences between the successive editions of a textbook being minimal, and 
as a result, the effective lifetime of these textbooks was extremely long. A 
noteworthy example is the Spanish translation of the work of A. Ganot --- 
the 1st edition was published in 1853 and the last we know of was 1945.

Likewise, there were difficulties in rooting out the antiquated content  of 
physics textbooks. In many of them, modern and old theories shared the 
pages. A significant example was the case of Eduardo S\'anchez Pardo, who 
amplified his translation \cite{Gan76} with elements of old editions that 
had been discarded from consideration in new foreign editions. The result 
was to help make the recovery of Spanish physics was a very slow process.

With respect to the term caloric, it was used in the Spain of this  period 
with different acceptations: as an imponderable fluid responsible for 
thermal phenomena (its original meaning), as a synonym for heat (the latter 
being understood according to the mechanical theory of heat), and as the 
cause of the phenomena of heat, whatever their nature.

Also, during the analysis of the sources that were available, we found  the 
orientation towards examination preparation to be excessive in the textbooks 
that we consulted. Some are a simple cookbook of laws and physical phenomena 
that the students would have to learn if they were to pass their 
examination. Hence, no interest was aroused in reflecting on content or in 
carrying out experiments in practical classes. Certain textbooks shamelessly 
encouraged totally memoristic learning, so that content was not applied to 
new situations and was readily forgotten.

Another  fact revealed by the analysis of the  
textbooks was the speed with which the mechanical theory of heat gained 
acceptance. This theory had become quite usual in many of the Spanish 
physics textbooks well before the concept of energy had begun to be treated 
in a general form. While this fact seems disconcerting from today's 
viewpoint, 
it must be pointed out that both the mechanical theory of heat and the second
law of thermodynamics had been proposed before the 
law of conservation of energy was generally accepted.

The introduction of the kinetic theory into 19th-century Spain occurred  
from approximately the 1870s onwards. In the 1880s, the theory already 
appears more developed in one of the textbooks that we consulted. This was 
the work of \citeauthor{Ami89} \shortcite{Ami89}, which stands out for its 
mechanistic deductions of Mariotte's Law and Avogadro's Law. The topics 
dealt with concerning the kinetic theory in the textbooks that we consulted 
were: generalities of the theory, kinetic interpretation of pressure, 
molecular velocities, mean free path, height of the atmosphere, and the van 
der Waals equation. Lastly it should be remarked that we found no original 
contribution to kinetic theory in these Spanish works of the end of the 
19th-century.
\section{Acknowledgements}
This work has been partially supported by the DGES (Spain) through Grant 
No.\ PB97-1501 and by the Junta de Extremadura (Fondo Social Europeo) 
through Grant No.\ IPR98C019.


\theendnotes

\end{document}